\documentclass[pre,aps,floatfix,showpacs,twocolumn]{revtex4}
\usepackage{graphicx}
\usepackage{bm}
\usepackage{amssymb}
\input{epsf}
\def\x{\left(\frac{r}{R}\right)}

\begin{document}

\title {Transition from mode-locked periodic orbit to chaos in a 2D
  piecewise smooth non-invertible map}


\author{Soma De} \email{soma@cts.iitkgp.ernet.in}
\affiliation{Department of Mathematics and Centre for Theoretical
  Studies,\\ Indian Institute of Technology, Kharagpur-721302, India}

\author{Soumitro Banerjee} \email{soumitro@ee.iitkgp.ernet.in}
\affiliation{Department of Electrical Engineering and Centre for
  Theoretical Studies,\\ Indian Institute of Technology,
  Kharagpur-721302, India}

\author{Akhil Ranjan Roy} \email{arroy@maths.iitkgp.ernet.in}
\affiliation{Department of Mathematics and Centre for Theoretical
  Studies,\\ Indian Institute of Technology, Kharagpur-721302, India${ }$}





\begin{abstract}

In this work we report a new route to chaos from a resonance torus in
a piecewise smooth non-invertible map of the plane into itself. The
closed invariant curve defining the resonance torus is formed by the
union of unstable manifolds of saddle cycle and the points of stable
cycle and saddle cycle. We have found that a cusp torus cannot develop
before the onset of chaos, though the loop torus appears. The
destruction of the two-dimensional torus occurs through homoclinic
bifurcation in the presence of an infinite number of loops on the
invariant curve. We show that owing to the non-invertible nature of the map,
the structure of the basin of attraction changes from simply connected
to a nonconnected one. We also describe how the mechanism of
transition to chaos differs from the scenario of appearance of chaos
in invertible maps as well as in smooth non-invertible maps.
\end{abstract}


\maketitle
  
\section{Introduction}

Piecewise smooth (PWS) maps have received much research attention in
the recent times because of their applicability in a large number of
practical systems
\cite{ban-book1,zhusubaliyev-book,bernardo_book}. This includes
switching circuits \cite{pre2d}, impact oscillators
\cite{brogl,Ing_sb08}, a variety of micro and macro economic systems
\cite{laugesen}, cardiac dynamics \cite{cm-berger-prl07} and many
other systems. For such maps a particular kind of bifurcation may
occur, which is different from what can occur in smooth maps. As a
control parameter changes, the fixed point may move in the phase space
and may collide with the borderline between two smooth regions,
resulting in an abrupt jump of the Floquet multipliers from the inside
of the unit circle to the outside of it in the complex plane, leading
to a special class of nonlinear phenomena known as border collision
bifurcation (BCB) \cite{Nusse92,Nusse94,pre2d,feigin-mario,Kowal}.

The BCB consists of many atypical phenomena such as direct transition
from a period-$1$ attractor to a chaotic attractor \cite{pre2d},
multiple attractor bifurcation \cite{mitrajit} and dangerous BCB
\cite{munther-prl}. Recently in a series of papers
\cite{somnath06,zhu08,zhu_ijbc08} a new type of BCB has been reported,
where a stable fixed point changes into an unstable focus along with
an attracting closed invariant curve. This closed invariant curve
forms a torus associated with quasiperiodic or phase-locked periodic
dynamics \cite{somnath06}. Moreover, the torus breakdown route to
chaos in PWS invertible map has been characterized in \cite{zhu08}
which is distinct from that of Afraimovich-Shilnikov
\cite{shilnikov}. These mechanisms of the creation and destruction of
tori were studied in the context of two-dimensional smooth invertible
maps \cite{aronson} and PWS invertible maps \cite{zhu08}.

In another line of work \cite{mais-mose,lorenz,frou}, the occurrence
of chaos via torus destruction has been studied numerically and
experimentally for non-invertible smooth maps. Several routes to chaos
were reported. The basic theory was proposed by Maistrenko {\em et
al.} \cite{mais-mose} and Lorenz \cite{lorenz}. In transition from a
smooth invariant curve (IC) to a chaotic attractor, the process begins
with the IC developing segments with increasingly high curvature. At a
certain parameter value cusps develop on the IC, which subsequently
change into loops. With further change of the parameter, the IC can be
destroyed in accordance with scenarios analogous to those of the
invertible case. After that the chaotic attractor comes into
existence. Cusps appear on IC when the degree of a map is greater than
one and the IC intersects the critical curve so that the tangent of
the IC at the point of intersection coincides with the eigenvector
corresponding to zero eigenvalue or the critical curve contains cusp
point (a point on a curve at which tangents of each branch of the
curve coincide). Critical curve is the two dimensional extension of
the local extrema (critical point) of a one dimensional differentiable
map. It is the locus of all those points at which Jacobian determinant
vanishes. For a non-differentiable piecewise linear map like the tent
map, the point of non-differentiability plays the role of critical
point. But for a two dimensional piecewise smooth map, according to
Mira {\em et al.} \cite{mira} the critical curve $LC$ is the image of
the borderline $LC_{-1}$ of the map and is characterized by the set
of all points which have at least two coincident rank-$1$
preimages. This curve may contain more than one segment. It separates
the phase plane into open regions where all points of a region have
the same number of rank-$1$ preimages.

The two dimensional PWS normal form maps \cite{pre2d,zhu08}
investigated so far are invertible. In the present paper we are
interested in the dynamical behavior of non-invertible PWS
maps. Considering a general piecewise linear 2D map expressed in the
normal form, we derive the parameter regions where the map is
non-invertible. We illustrate how chaos appears in PWS non-invertible
maps and demonstrate the difference from the corresponding mechanisms
of transition for smooth invertible \cite{aronson} and non-invertible
maps \cite{mais-mose,lorenz}. We investigate the bifurcation that
takes place as one moves from the inside of a resonance tongue to the
outside in the parameter space. Inside a resonance tongue, a closed
invariant curve is formed by the saddle-node connections of a pair of
cycles (a saddle and a node). This curve forms the mode-locked
torus. We show that the cusp torus does not develop but the loop torus
\cite{mais-mose} is created before the destruction of mode-locked
torus as the parameter is varied continuously. The mode-locked torus
is destroyed through a homoclinic bifurcation. We observe that the
interaction between stable manifold and the critical curve leads to a
qualitative change in the basin of attraction, from a simply connected
to a nonconnected one. We also show that the unstable manifolds can
have structurally stable self-intersections while the stable manifolds
can intersect itself when the map becomes structurally unstable. There
is a coexistence of stable node and a chaotic attractor. We show that
the size of basin of attraction of the stable cycle (node) diminishes
to zero and after this the stable node undergoes a border collision
fold bifurcation. We also illustrate the important role played by
non-invertibility of the map in the mechanism of transition to chaos.

\section{Considered system}
For a piecewise smooth map whose leading-order Taylor term close to
the border is linear, the dynamics in the neighborhood of the border
can be expressed by a normal form map \cite{pre2d,Nusse92} defined by

\begin{equation} 
 F : \left ( \begin{array}{c} x \\ y \end{array}\right) \rightarrow
\left\{\begin{array}{cc} F_1 (x, y) & \mbox{if}\;\;
x \leq 0\\ F_2 (x, y) & \mbox{if}\;\; x \geq 0
\end{array} \right. \label{2dmap}
\end{equation}

where
\[
F_1 (x, y) = \left(\begin{array}{c} \tau_L x +y+\mu \\ -\delta_Lx
\end{array}\right),\]
\[ F_2 (x, y) = \left(\begin{array}{c} 
\tau_R x +y+\mu \\ -\delta_Rx
\end{array}\right),\;\; (x, y) \in \mathbb{R}^2.
\] 

The phase plane $\mathbb{R}^2$ is divided into two compartments
$L:=\{(x,y)\in \mathbb{R}^2:x \leq 0\}$ and $R:=\{(x,y)\in
\mathbb{R}^2:x >0\}$. $\tau_L$ and $\delta_L$ are the trace and
determinant respectively of the Jacobian matrix $J_L$ in the left
half-plane $L$, and $\tau_R$ and $\delta_R$ are the trace and
determinant respectively of the Jacobian matrix $J_R$ in the right
half-plane $R$, and $\mu$ is a parameter. Thus
\[J_L= \left( \begin{array}{cc} \tau_L & 1 \\ -\delta_L &
0 \end{array}\right) \;\; \mbox{and}\;\; J_R = \left(
\begin{array}{cc} \tau_R & 1 \\ -\delta_R & 0 \end{array}\right).\]

Most of the theories of BCB discussed so far are on dissipative
dynamical systems (i.e., $|\delta_L| < 1$, $|\delta_R| < 1$). In
contrast, we consider a situation where the map is contracting in the
left side and expanding in the right side of the border, which happens
when $|\delta_L|< 1$, $|\delta_R|> 1$. We choose $\tau_L,\; \tau_R$
satisfying the conditions
\begin{equation}
-(1+\delta_L)<\tau_L<(1+\delta_L), \;\;\;
 -2\sqrt{\delta_R}<\tau_R<2\sqrt{\delta_R}.\label{preg}
\end{equation}

\begin{figure}

\centerline{\epsfxsize=2.8in\epsffile{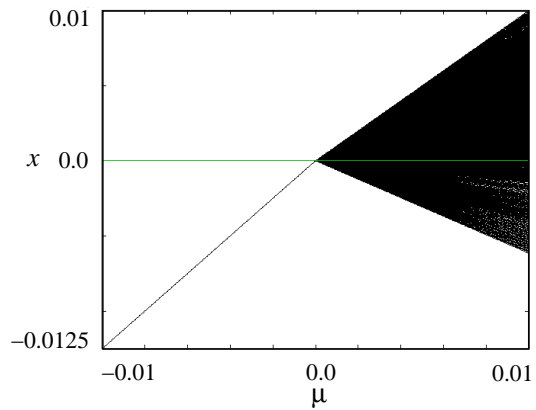}}

\caption{\label{1stfig} Bifurcation diagram obtained by varying the
  parameter $\mu$ from $-0.01$ to $0.01$. This diagram illustrates the
  birth of a quasiperiodic orbit after the border collision, directly
  from a stable fixed point for $\delta_L=-0.3$, $\delta_R=1.4$,
  $\tau_L=-0.1$ and $\tau_R=-0.685$.}

\end{figure}


The above conditions ensure that the fixed point for $\mu<0$ is
attracting and for $\mu>0$ it is a spiral repellor. According to the
notations in \cite{feigin-mario}, we can write the stable fixed point
by $A$ and unstable fixed point by $a$ in $L$, and similarly by $B$
and $b$ respectively in $R$. The fixed point in $L$ is
\[
A \;or\; a = (\frac{\mu}{1-\tau_L+\delta_L},
\frac{-\mu\delta_L}{1-\tau_L+\delta_L})
\]
and the fixed point in $R$ is
\[
B \;or\; b = (\frac{\mu}{1-\tau_R+\delta_R},
\frac{-\mu\delta_R}{1-\tau_R+\delta_R}).
\]
$A$ or $a$ exists if ${\mu}/{(1-\tau_L+\delta_L)} \leq 0$, otherwise a
virtual fixed point is located in $R$ and is denoted by $\bar{A}$ or
$\bar{a}$. Similarly $B$ or $b$ exists if ${\mu}/{(1-\tau_R+\delta_R)}
\geq 0$, otherwise a virtual fixed point is located in the $L$ side
and is denoted by $\bar{B}$ or $\bar{b}$. The stability of the fixed
points is given by the eigenvalues
\[
\lambda_1, \lambda_2 = \frac{1}{2}(\tau\pm \sqrt{\tau^2-4\delta}).
\] 
As the parameter $\mu$ hits the border at $\mu=0$, a stable fixed
point $A$ on the $L$ side becomes an unstable focus $b$ on the $R$
side. When the outward spiraling orbit in the $R$ side touches the
boundary of the $L$ side then it is attracted by the virtual fixed point
$\bar{A}$ located in $R$. So the outward motion is arrested. This gives
rise to the existence of an invariant closed curve on which stationary
state must lie. Fig.~\ref{1stfig} displays the bifurcation diagram
varying the parameter $\mu$ from a negative to a positive value. This
diagram depicts an abrupt transition from periodic to quasiperiodic
behavior at a border-collision bifurcation.

\begin{figure}

\centering 
\includegraphics[width=.95\columnwidth]{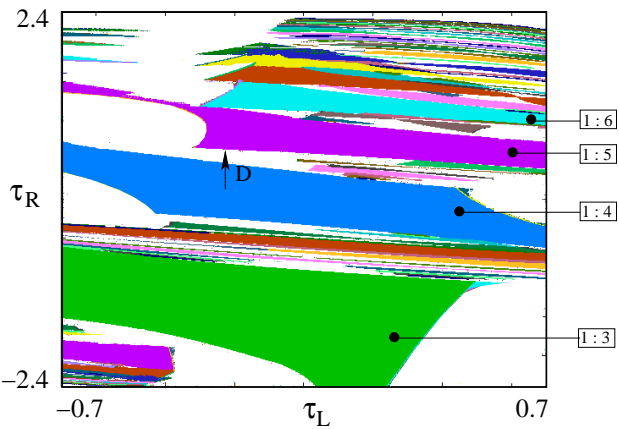}
\caption{\label{2ndfig} Regions of the periodic tongues in the
  ($\tau_L, \tau_R$) parameter plane for $\delta_L=-0.3$,
  $\delta_R=1.4$ and $\mu=0.05$. Distinct colors of the tongues denote
  different oscillatory motions.}

\end{figure}

We plot the different resonance tongues in ($\tau_L, \tau_R$)
parameter plane for a positive value of $\mu$ (see
Fig.~\ref{2ndfig}). The main resonance tongues are denoted with the
corresponding rotation numbers. Inspection of this figure reveals the
presence of different periodic tongues which are densely spread in the
parameter space. Yang and Hao \cite{yang-hao} described the lens-like
resonance tongues for a one-dimensional piecewise linear system, which
were later generalized for two and three dimensional piecewise smooth
systems in \cite{zhusubaliyev-book,zhusubaliyev02,simpson}. The
tongues are bounded by the curves where the stable and unstable cycles
merge and disappear through BCB. A tongue may contain more than one
lens.

\section{Properties of the considered non-invertible map}

\medskip
\noindent {\bf Proposition $1$.} { \it The two dimensional piecewise
  smooth normal form map (\ref{2dmap}) is non-invertible, if either
  $\delta_L$ and $\delta_R$ are of opposite sign, or any one of
  $\delta_L$ and $\delta_R$ is zero.}

\vspace{0.2cm}

\noindent {\bf Proof.} Since our map is continuous in $\mathbb{R}^2$,
so we have to derive the conditions from one to one property. Let us
suppose

\[X_1=\left (
\begin{array}{c} x_1 \\ y_1 \end{array}\right), \;X_2=\left (
\begin{array}{c} x_2 \\ y_2
\end{array}\right)\; \mbox{and}\; X_1,\; X_2 \in\mathbb{R}^2.
\]

 There are three possibilities.
\vspace{.1cm}

\noindent {\bf Case I:} $X_1$, $X_2$ $\in{L}$.

Now $ F_1(X_1)=F_1(X_2)$
\newline$\Rightarrow \tau_Lx_1+y_1+\mu=\tau_Lx_2+y_2+\mu\;$
$\mbox{and}\; -\delta_Lx_1=-\delta_Lx_2$.
\newline $\Rightarrow x_1=x_2 \;$(if$\;\delta_L\neq0$)
$\mbox{and} \;y_1=y_2$ .
\newline So $F_1$ is one-one if $\delta_L\neq0$.

\vspace{.1cm}

\noindent {\bf Case II:} $X_1$, $X_2$ $\in{R}$.

By a similar argument as in Case I, we conclude that $F_2$ is one-one
if $\delta_R\neq0$.

\vspace{.1cm}

\noindent {\bf Case III:} $X_1$ $\in{L}$, $X_2$ $\in{R}$.
\newline$ F_1(X_1)=F_2(X_2)$
\newline$\Rightarrow \tau_Lx_1+y_1+\mu=\tau_Rx_2+y_2+\mu\;$
$\mbox{and}\; -\delta_Lx_1=-\delta_Rx_2$.
\newline $\Rightarrow x_1=(\delta_R/\delta_L)x_2\;$(if $\;\delta_L\neq0$). 

\noindent (a) If $\delta_L$ and $\delta_R$ are of the same sign and
are non-zero.

Since $x_1\leq 0$ and $x_2>0$, the relation
$x_1=(\delta_R/\delta_L)x_2\;$ cannot hold. Therefore there can exist
no points $X_1 \in {L}$, $X_2 \in {R}$ such that
$F_1(X_1)=F_2(X_2)$. Hence the map is invertible.

\noindent (b) If $\delta_L$ and $\delta_R$ are of opposite sign.

The condition $x_1=(\delta_R/\delta_L)x_2$ may hold.

Then $ F_1(X_1)=F_2(X_2)$ $\Rightarrow {X_1\neq X_2}$ because $x_1\leq
0$ and $x_2>0$. Hence the map is non-invertible. Combining the above
three cases the map is non-invertible if one of the following
relations is satisfied

\noindent $(i)$ $\delta_L$ and $\delta_R$ are of opposite sign;

\noindent $(ii)$ any one of $\delta_L$ and $\delta_R$ is zero. $\;\; \Box$

\vspace{0.2cm}

To consider a typical non-invertible case, we assume $\delta_L=-0.3$
and $\delta_R=1.4$. Each point in $\mathbb{R}^2$ is mapped under
(\ref{2dmap}) to a point in $\mathbb{R}^2$ with $y$-coordinate less than
or equal to zero. There are two inverses for $(x,y)$ in $\mathbb{R}^2$
with $y\leq 0$. These are defined as

\[F^{-1}_1 (x, y) = \left(\begin{array}{c}(-1/{\delta_L})y\\
x+(\tau_L/\delta_L)y-\mu\end{array}\right) \; \mbox{and}\]
\[
F^{-1}_2 (x, y) = \left(\begin{array}{c}(-1/{\delta_R})y
\\x+(\tau_R/\delta_R)y-\mu\end{array}\right),
\]
where $(x, y) \in \mathbb{R}^2$ and $y\leq 0$. 

Hence a point in $\mathbb{R}^2$ has two distinct rank-$1$ preimages if
$y<0$, no preimage if $y>0$ and two coincident preimages if $y=0$. Let
$Z_0$ be the region with no preimage (the upper half plane) and $Z_2$
be the region with two different preimages (the lower half
plane). Thus for our choice of parameters, the map is of $Z_0-Z_2$
type \cite{mira}. The locus of the points having two merging rank-$1$
preimages is the $x$-axis, called $LC$. Here $LC$ is the critical curve.
The preimages of each point in $LC$ lie on the $y$-axis, named as
$LC_{-1}$. So the non-invertible map folds the phase plane along
$LC$. The fixed points, periodic orbits and attractor must lie inside
$Z_2$, because every fixed point and every point of a periodic orbit
have an inverse (namely itself). A point on an attractor also has a
preimage lying on that attractor.

\section{Structure of the Stable and Unstable manifolds}

Let $f:\mathbb{R}^2 \rightarrow \mathbb{R}^2$ be a non-invertible
smooth map and $x$ be a hyperbolic period-$q$ point of $f$. We define
the stable manifold of $x$ as $W^s(x)=\{y\in\mathbb{R}^2:
f^{nq}(y)\rightarrow x \;\mbox{as}\ n \rightarrow \infty \}$ and the
unstable manifold of $x$ as
$W^u(x)=\{y\in\mathbb{R}^2:f^{-nq}(y)\rightarrow x \;\mbox{as}\ n
\rightarrow \infty \}$. These are defined in the neighborhood of $x$,
and hence so are actually the local stable and unstable manifolds of
$x$ \cite{robinsonbook}. The global unstable manifold is formed by the
union of forward images of the local unstable manifold. It is
connected, since the local unstable manifold is connected and the
image of a connected set under continuous map is connected. The global
unstable manifold is forward invariant but may not be backward
invariant (some of its points have multiple preimages) and so
self-intersections are allowed. The global stable manifold is formed
by the union of all preimages of any rank of the local stable
manifold. Since the map $f$ is non-invertible, the global stable
manifold is backward invariant, but it may be strictly mapped into
itself (some of its points may have no preimage). So it may be
disconnected. For non-invertible maps the use of the term ``manifold''
is an abuse of the terminology; the local stable and unstable
manifolds are guaranteed to be true manifolds, but the global
manifolds are not \cite{frouzakis97, robinsonbook}. Still for our map
we denote the global stable and unstable manifolds as the stable and
unstable manifolds respectively.

Since our map $F$ is of $Z_0-Z_2$ type and the local unstable manifold
originates at the saddle point $x$, it must lie in the $Z_2$
region. The unstable manifold is grown by iterating the local unstable
manifold forward in time, every point computed will have a preimage,
namely the previous point iterated to arrive at that point and hence
the unstable manifold will never leave the $Z_2$ region (see
Fig.~\ref{3rdfig}(a)). The stable manifold is the union of all
preimages of local stable manifold, connected or disjoint, so it can
lie inside the $Z_0$ region also (see Fig.~\ref{3rdfig}(a)).

\section{Transition to chaos in the presence of homoclinic structure}

\begin{figure}[tbh] 
\centering 
\includegraphics[width=.65\columnwidth]{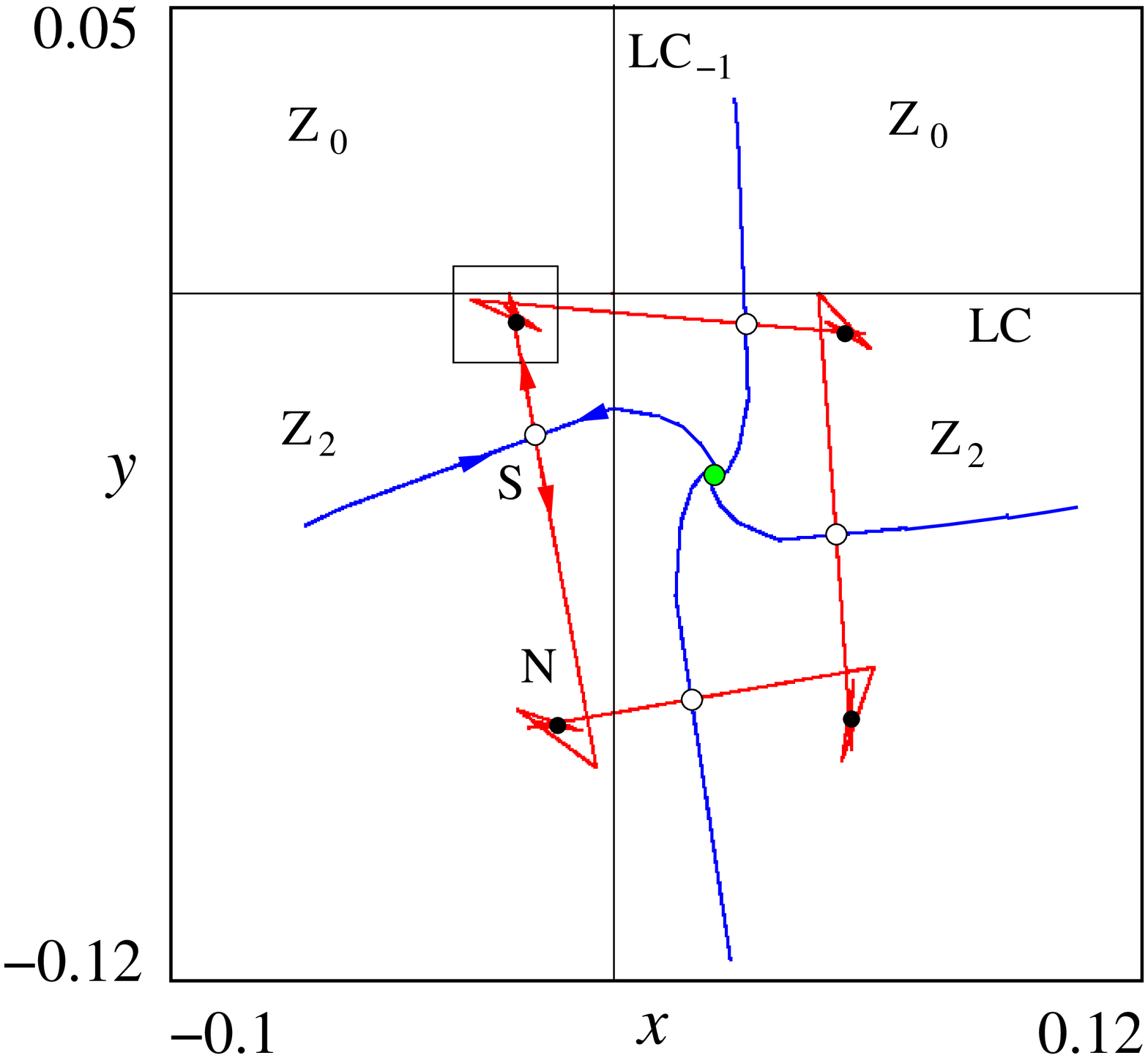}{\small
(a)} \hspace{1.5cm}
\includegraphics[width=.65\columnwidth]{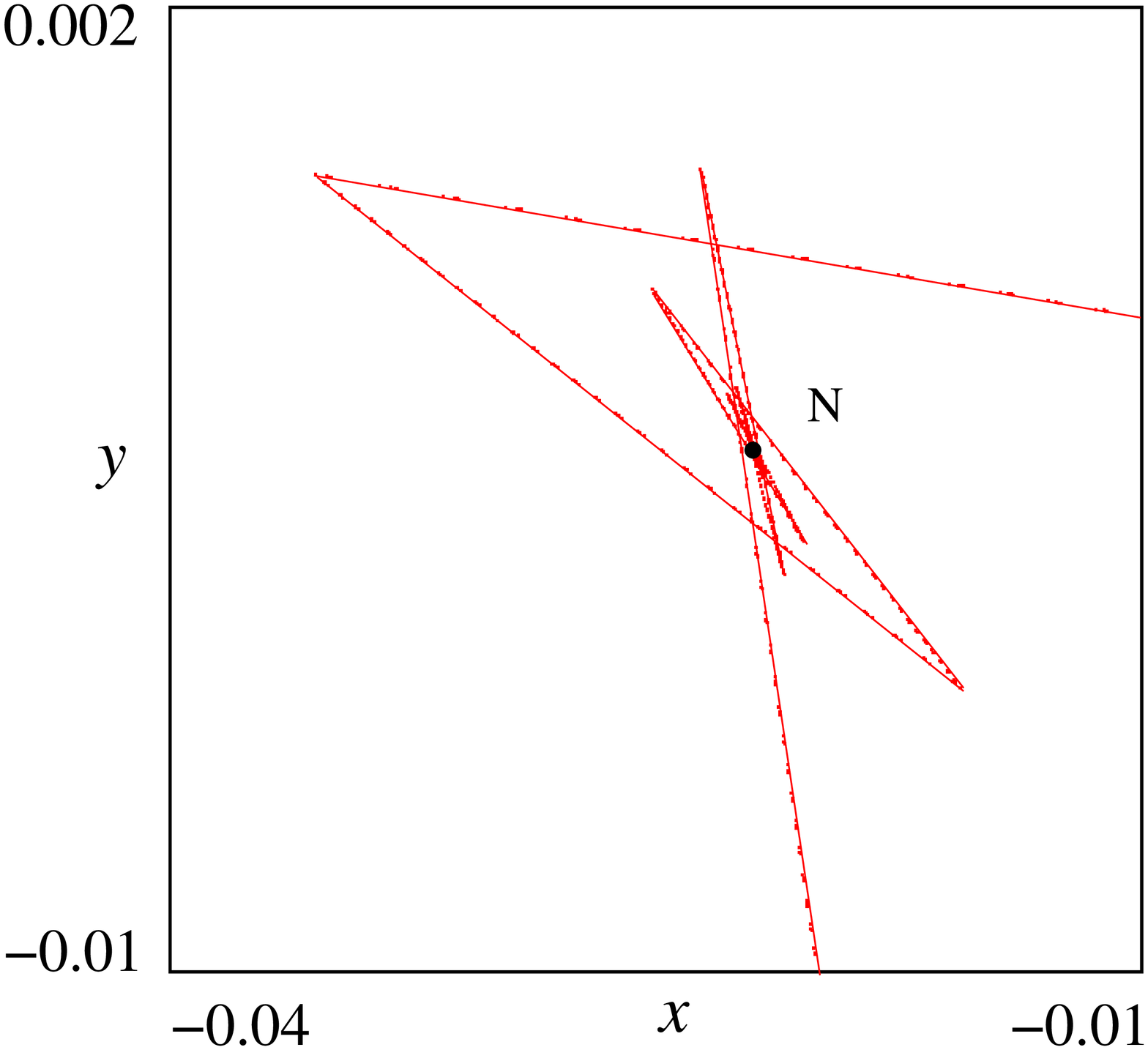}{\small
(b)}
\caption{\label{3rdfig} (a) Phase portrait of the attractor at
  $\tau_R=0.25$ with $\tau_L=-0.3$, $\delta_L=-0.3$, $\delta_R=1.4$
  and $\mu=0.05$. (b) Enlargement of a portion of the IC marked by the
  square in (a). Points of stable cycles are marked with solid black
  circles, points of saddle cycles with open circles and the spiral
  repellor is marked with solid green circle.}

\end{figure}

\begin{figure}
\centering 
\includegraphics[width=.8\columnwidth]{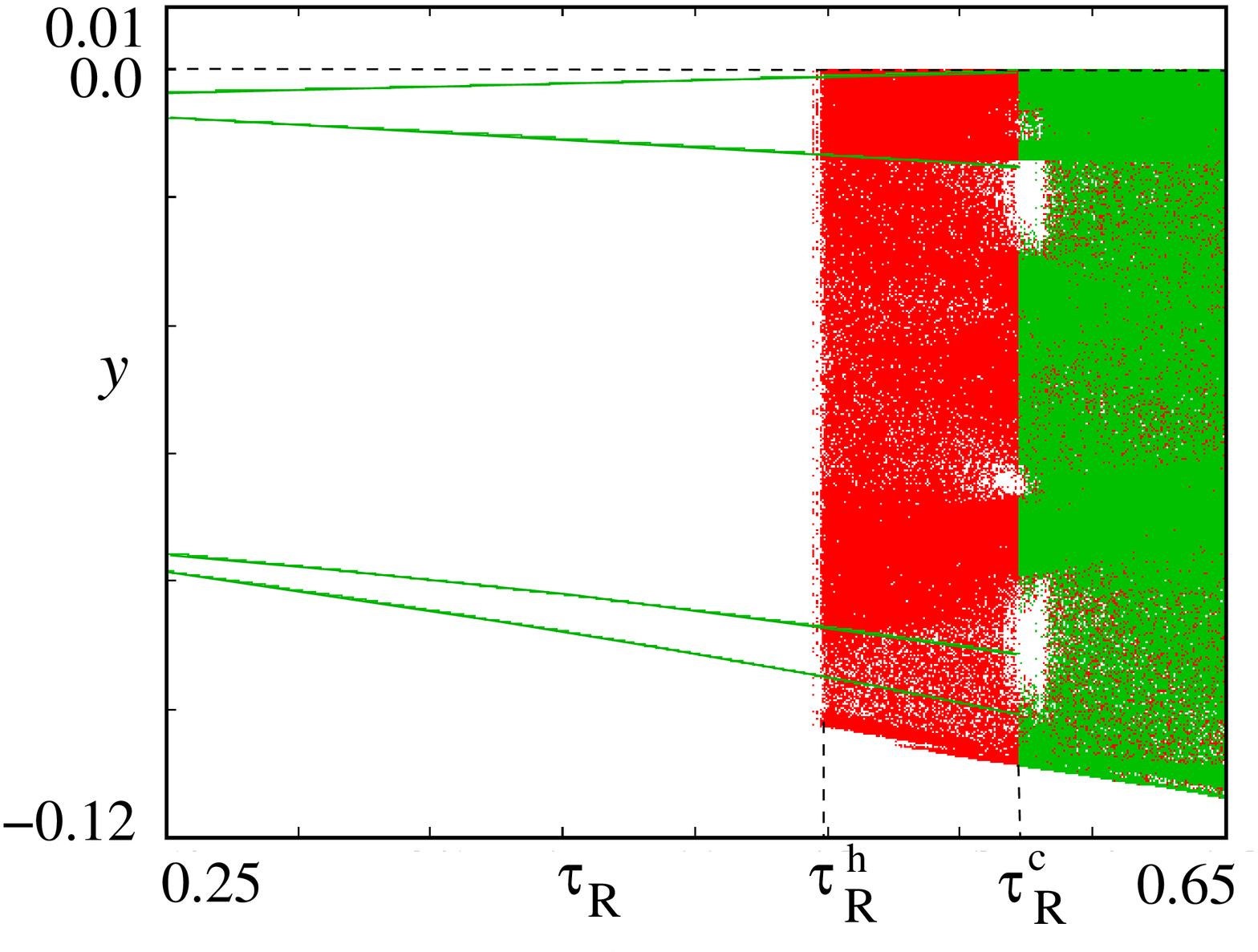}{\small (a)}
\hspace{1cm}
\includegraphics[width=.78\columnwidth]{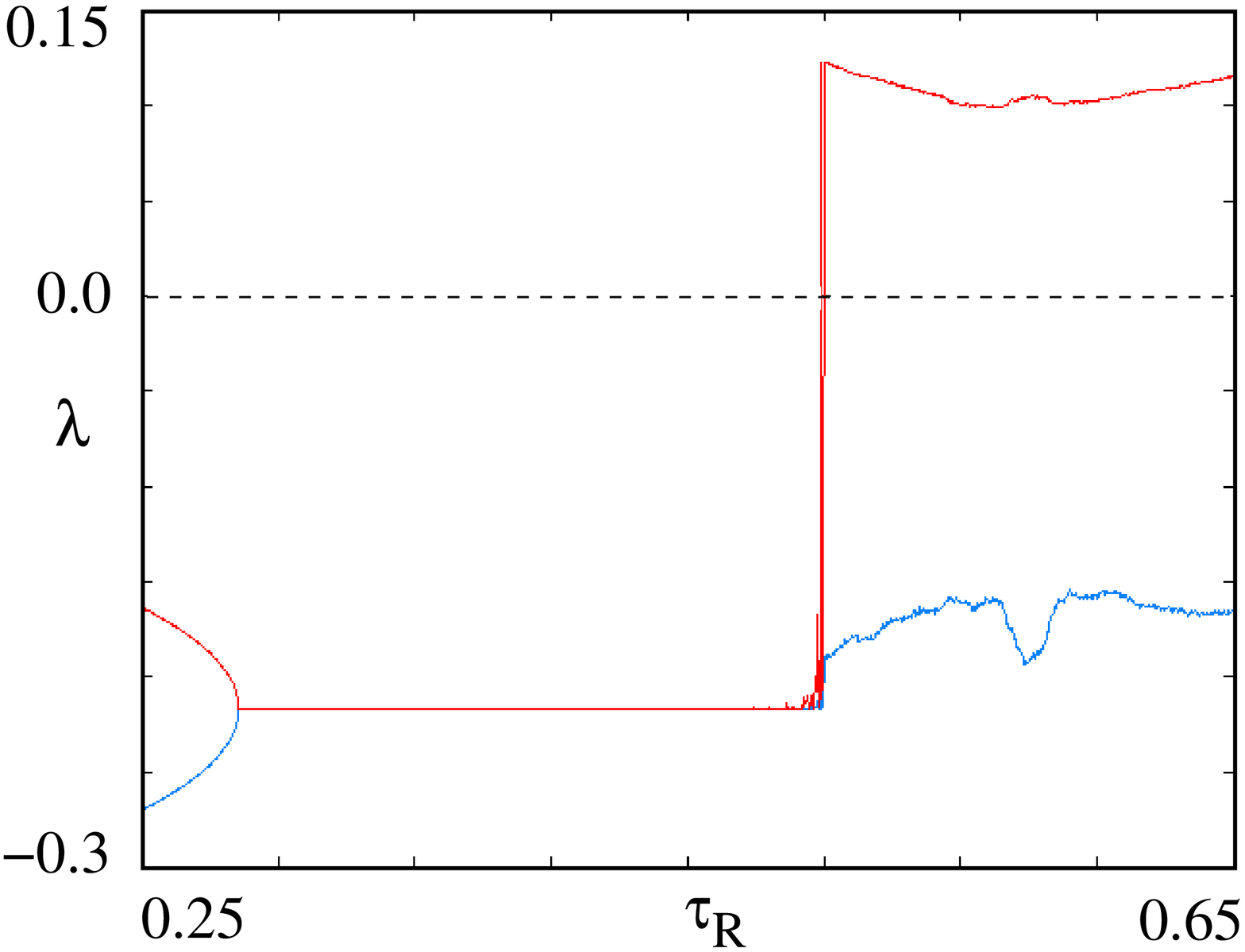}{\small(b)}
\caption{\label{4thfig}(a) Bifurcation diagram illustrating the birth
  of a chaotic orbit via border-collision saddle-focus bifurcation and
  (b) the corresponding Lyapunov exponents. The bifurcation parameter
  $\tau_R$ changes from $0.25$ to $0.65$, for $\tau_L=-0.3$,
  $\delta_L=-0.3$, $\delta_R=1.4$ and $\mu=0.05$. The period-$4$ orbit
  coexists with the chaotic orbit in the region $\tau_R^h < \tau_R <
  \tau_R^c$, where $\tau_R^h \approx 0.5$ and $\tau_R^c \approx
  0.57414$.}
\end{figure}

Let us now analyze the bifurcational transition from mode-locking to
chaos. To illustrate this scenario, we vary the bifurcation parameter
$\tau_R$ from the inside of the $1:4$ tongue to the outside of it
along the direction $D$ (see Fig.~\ref{2ndfig}). The reason for this
choice is that, the transition to chaos maintains the loop like
structure as described by Maistrenko {\em et al.} \cite{mais-mose}
which we want to explain. Fig.~\ref{4thfig}(a) displays the
bifurcation diagram and Fig.~\ref{4thfig}(b) depicts the Lyapunov
exponents corresponding to the bifurcation diagram. The largest
Lyapunov exponent becomes positive at $\tau_R\approx 0.5$, which
implies the existence of a chaotic attractor. The second Lyapunov
exponent remains negative over the considered parameter
range. Initially the map (\ref{2dmap}) has a period-$4$ stable cycle
and a period-$4$ saddle cycle. As the parameter $\tau_R$ increases,
between $\tau^h_R\approx 0.5$ and $\tau^c_R\approx 0.57414$ the stable
period-$4$ cycle coexists with a chaotic attractor. So we observe hard
hysteretic transitions from periodic mode to chaotic mode and vice
versa, at the points $\tau^h_R$ and $\tau^c_R$ respectively. At
$\tau^c_R$ one of the points of the period-$4$ stable cycle and a
point of the period-$4$ saddle cycle hit the border together. Then the
period-$4$ stable and saddle cycles disappear through border collision
fold bifurcation.
 
%
%
%
\begin{figure}
\centering
\includegraphics[width=.65\columnwidth]{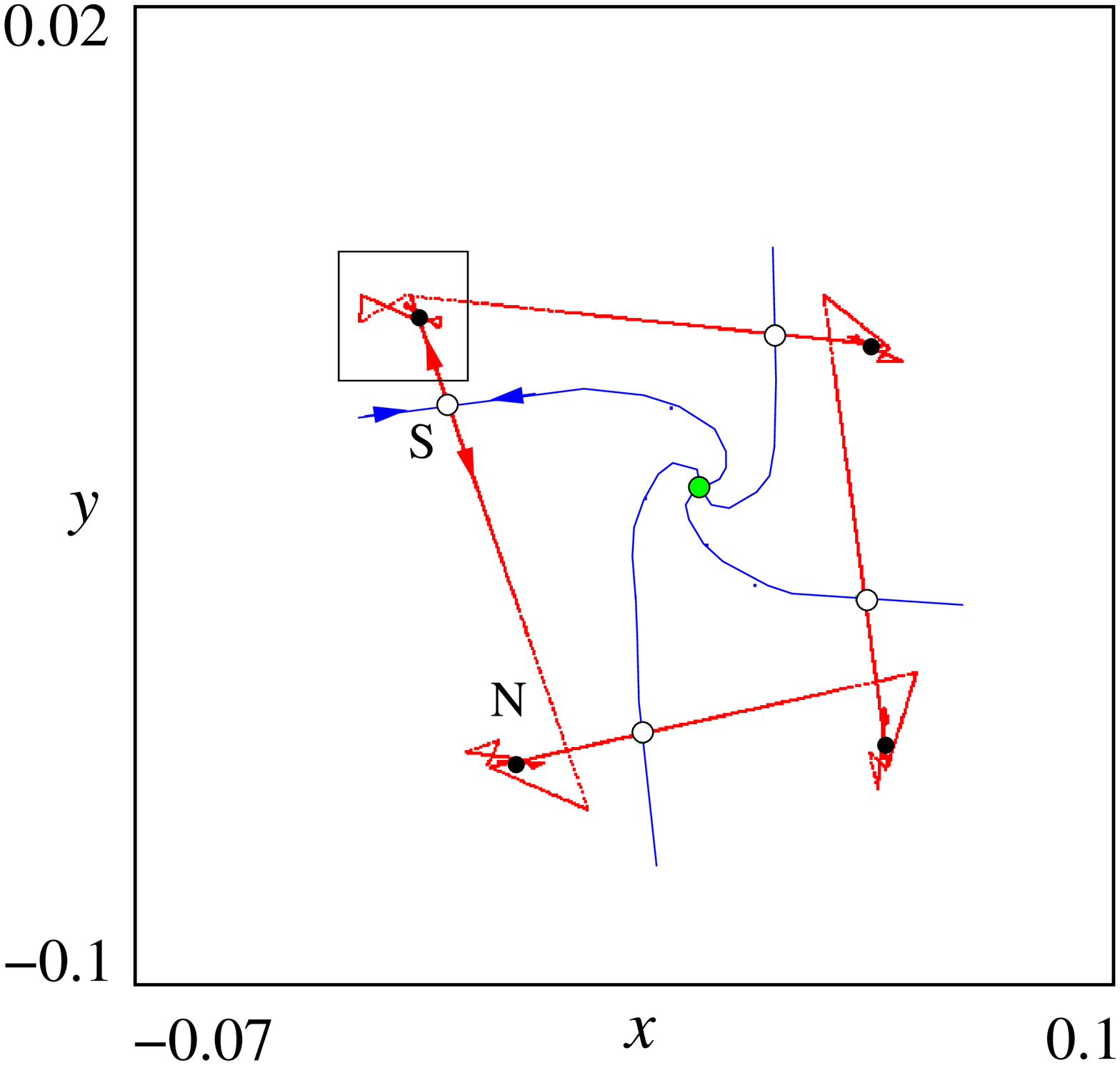}{\small (a)}
\hspace{1.5cm}
\includegraphics[width=.65\columnwidth]{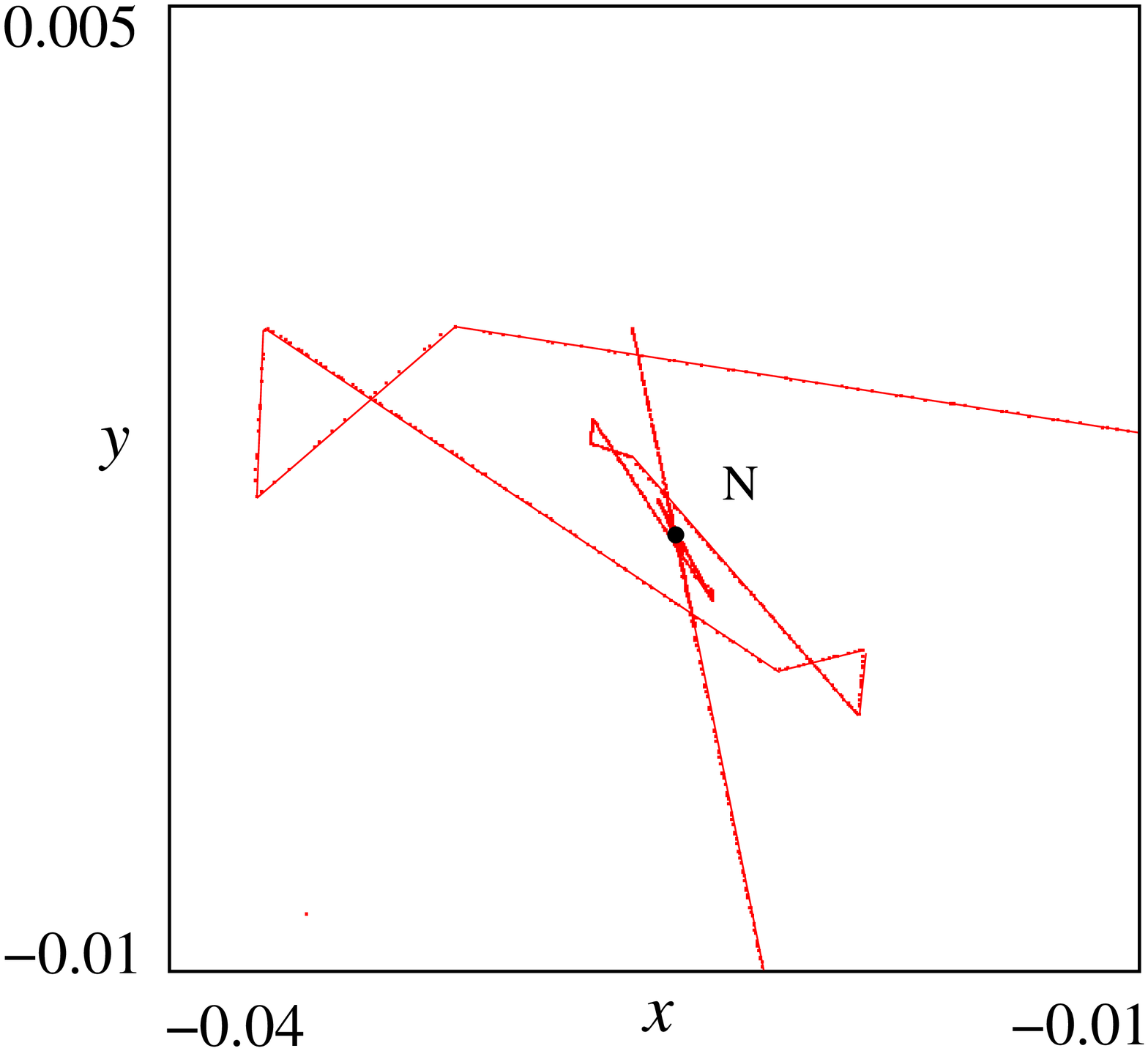}{\small (b)}
\caption{\label{5thfig} (a) Phase portrait of the attractor at
  $\tau_R=0.28$ with $\tau_L=-0.3$, $\delta_L=-0.3$, $\delta_R=1.4$
  and $\mu=0.05$. (b) Magnified part of the IC marked by the square in
  (a) displays the loops on IC.}

\end{figure}
\begin{figure}[tbh] 
\centering
\includegraphics[width=.65\columnwidth]{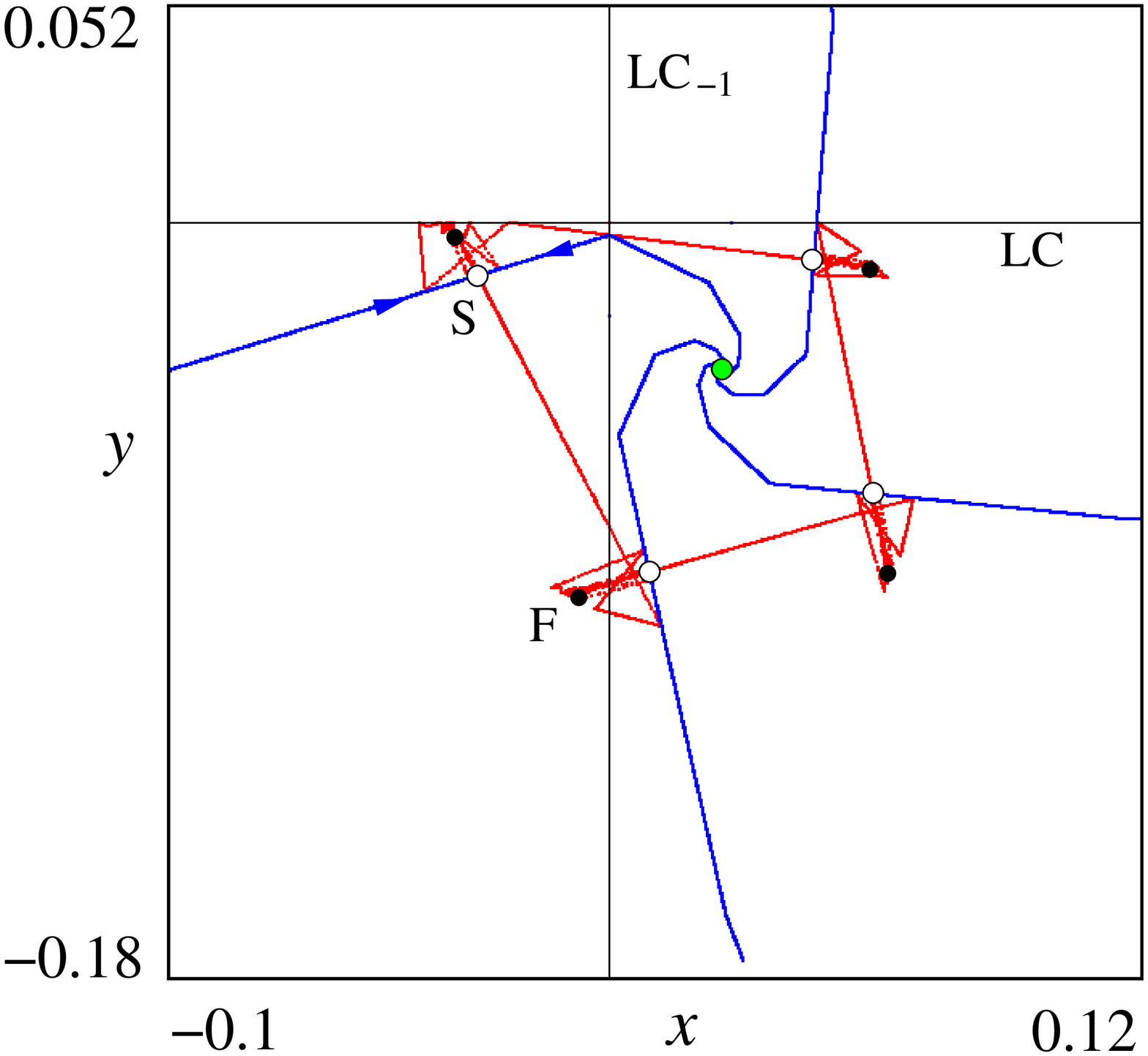}{\small (a)}
\hspace{1.5cm}
\includegraphics[width=.65\columnwidth]{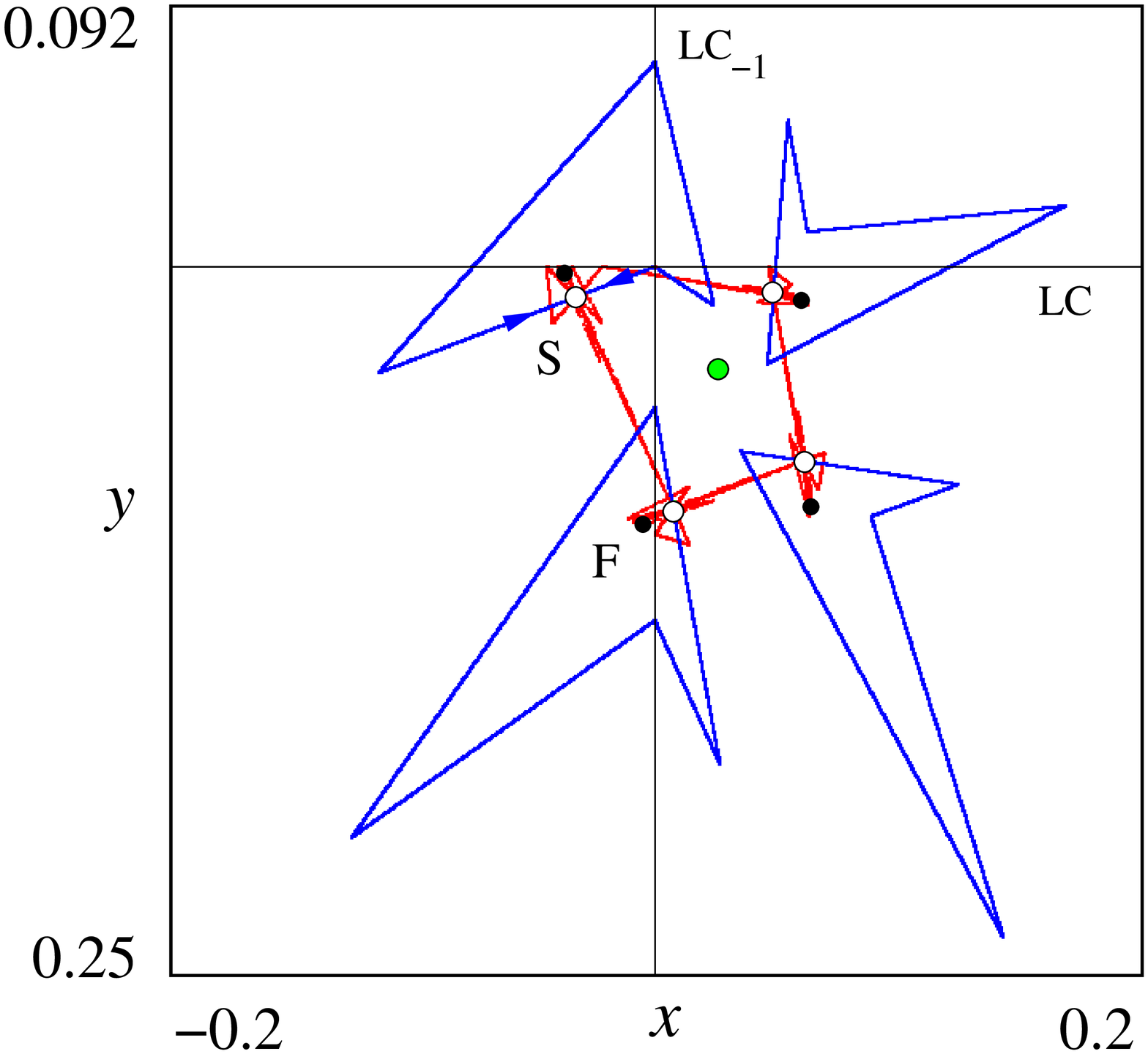}{\small (b)}

\caption{\label{6thfig} Phase portraits of the attractors at (a)
  $\tau_R=0.4$ and (b) $\tau_R=0.4295$ with other parameters same
  as in Fig.~\ref{3rdfig}.}

\end{figure}

\begin{figure}[tbh]
\centering
\includegraphics[width=.65\columnwidth]{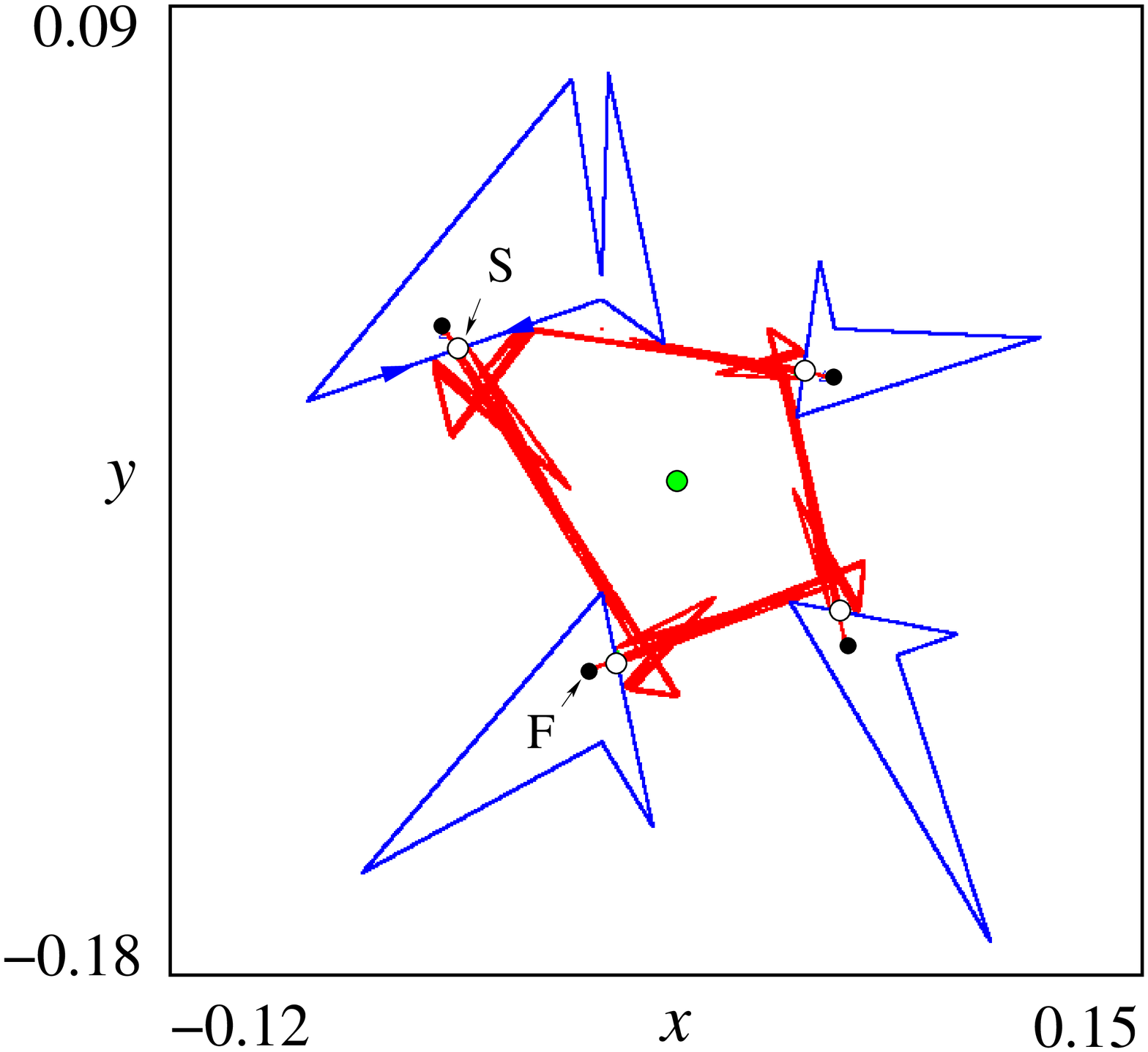}{\small (a)}
\hspace{1.5cm}
\includegraphics[width=.65\columnwidth]{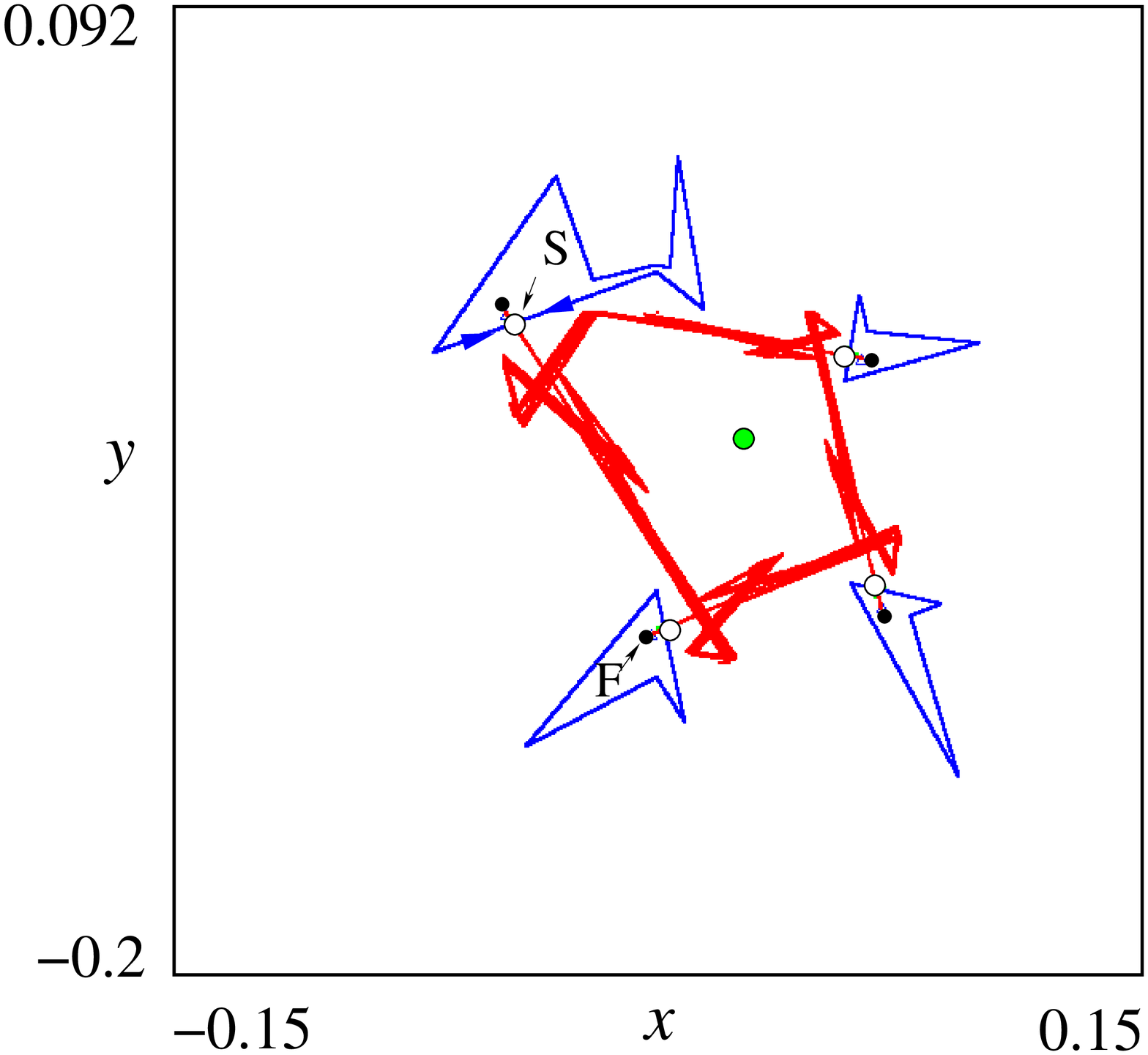}{\small (b)}
\caption{\label{7thfig} Attractors are depicted at (a) $\tau_R=0.5$
  and (b) $\tau_R=0.53$ with other parameters same as in
  Fig.~\ref{3rdfig}.}

\end{figure}

Let us now characterize the mechanism of occurrence of chaos in
detail. The phase portrait at $\tau_R=0.25$ is shown in
Fig.~\ref{3rdfig}(a). The period-$4$ attracting cycle has real
eigenvalues. The period-$4$ saddle cycle always has one positive
eigenvalue $\lambda_1>1$ and another eigenvalue
$-1<\lambda_2<0$. Since the stable eigenvalue of the saddle cycle is
negative, so our map is orientation reversing and hence upon
successive iterations, points on the stable manifold oscillate about
the saddle point as they approaches to it. The closed invariant curve
is formed by the union of the unstable manifolds of period-$4$ flip
saddle cycle and the points of the period-$4$ regular attractor and
flip saddle cycles. It is non-smooth, since the unstable manifolds are
as smooth as the original map and our map is piecewise smooth. This
curve forms the mode-locked torus. Fig.~\ref{3rdfig}(a) shows that the
unstable manifolds are self-intersecting. As our map is an
endomorphism, the unstable manifold of a saddle point can intersect
itself as well as the unstable manifold of another saddle point of the
same saddle cycle. At $\tau_R=0.25$ the unstable manifold of a saddle
point intersects the unstable manifold of another saddle point of the
same period-$4$ saddle cycle (see Fig.~\ref{3rdfig}(b)). Both types of
self-intersections appear at $\tau_R\approx 0.28$ (see
Fig.~\ref{5thfig}(a)). Since our map is non-invertible, a point on the
unstable manifold can have two distinct preimages so the unstable
manifolds are self-intersecting, though they are connected. From the
definition of unstable manifold, we can say that there are infinite
number of self-intersections (for both types) because the image of a
self-intersecting point is again a self-intersecting point. With
increase of the parameter, at $\tau_R\approx 0.28$ loops are created
on the invariant curve (see Fig.~\ref{5thfig}(a)). Since at the points
of self-intersections (first type) loops are developed so the infinite
loop structure is formed on the invariant curve (see
Fig.~\ref{5thfig}(b)). The torus formed by this invariant curve is a
loop torus \cite{mais-mose}. As the parameter is increased to
$\tau_R\approx 0.2855$, the eigenvalues of the stable cycle become
complex conjugate. Then the closed invariant curve is formed by the
union of the unstable manifolds of period-$4$ flip saddle cycle and
the points of the period-$4$ spiral attractor and flip saddle
cycles. Again increasing $\tau_R$, the unstable manifolds of the
period-$4$ flip saddle cycle nontransversally touch the stable
manifolds of it. Then the first homoclinic tangency appears at
$\tau_R\approx 0.4$ (see Fig.~\ref{6thfig}(a)). Since the stable
eigenvalue of the saddle cycle is negative, the homoclinic orbit flips
around the saddle point. Thus the homoclinic tangency occurs in both
sides of the stable manifold. With further increase of $\tau_R$, the
stable and unstable manifolds of the period-$4$ flip saddle cycle
intersect transversally to form a homoclinic tangle (see
Fig.~\ref{6thfig}(b)). The existence of homoclinic tangle gives rise
to Smale horseshoe dynamics, consequently an infinite number of high
periodic orbits are created. The torus no longer exists, but the
period-$4$ spiral attractor and the flip saddle cycle persist. Further
increasing $\tau_R$, the second homoclinic tangency occurs at
$\tau^h_R\approx 0.5$ (see Fig.~\ref{7thfig}(a)). The period-$4$
spiral attractor coexists with a chaotic attractor between the
parameter values $\tau^h_R$ and $\tau^c_R$. At $\tau^c_R\approx
0.57414$ the period-$4$ spiral attractor and flip saddle cycles merge
and disappear through border collision fold (saddle-focus)
bifurcation.
%
This scenario of transition to chaos is quite different from that in
invertible maps since the self-intersections of the unstable manifolds
and creation of loops on the invariant curve are impossible in
invertible maps \cite{aronson, zhu08}. There is only one attractor
before the first homoclinic tangency, namely, the closed invariant
curve consisting of the periodic orbits. After the appearance of the
homoclinic tangle, the closed invariant curve is destroyed, yet the
period-$4$ spiral attractor continues to exist. The stable manifold of
the saddle point to the left side of the border ($x=0$) touches the
critical curve LC at $\tau_R \approx 0.4295$ (see
Fig.~\ref{6thfig}(b)). Then a contact bifurcation \cite{mira}
occurs. It causes changes in the basin of attraction of the period-$4$
spiral attractor, simply connected to nonconnected and unbounded to
bounded, since the basin boundary is formed by the stable manifolds of
the saddle cycle. A different basin of attraction for the period-$4$
attractor is created. The stable manifolds of each of the four saddle
points are disconnected from each other. The stable manifold of a
non-invertible map can intersect itself when the map is structurally
unstable. Owing to the presence of the homoclinic tangle, our map has
already become structurally unstable. So the stable manifolds of the
period-$4$ flip saddle cycle form nonconnected (with finite number of
simply connected components) and bounded basin of attraction of the
period-$4$ spiral attractor. To our knowledge, this phenomenon in PWS
dynamics has not been reported earlier. The stable manifolds fold at
every intersection with the $y$-axis \cite{pre2d}. Other fold points
on the stable manifolds are preimages of the fold points on the
$y$-axis in $Z_2$. As we increase $\tau_R$, the size of the basin of
attraction of the period-$4$ spiral attractor diminishes (see
Fig.~\ref{7thfig}(b)).  This size tends to zero just before the border
collision fold bifurcation.

\section{Conclusion}

The study of non-invertible dynamical systems is an important and
recently flourishing research subject. In contrast to the earlier
investigations \cite{pre2d,somnath06,zhu08,zhu_ijbc08}, here we have
chosen the situation where the map (\ref{2dmap}) is non-invertible,
and the determinant of the Jacobian matrix in one side of the
bifurcation point is less than unity and is greater than unity in the
other side. Such a situation arises in many physical systems like
power electronic circuits. In this paper we have illustrated a
scenario of transition from phase-locked dynamics to chaos in such a
non-invertible piecewise smooth map. The closed invariant curve is
formed by the union of the unstable manifolds of a saddle cycle and
the points of stable and saddle cycles. Since a point can have two
distinct preimages, the unstable manifolds have structurally stable
self-intersections. We have found that loops are created on the
invariant curve although there is no chance of appearance of cusps on
it. As our map is linear in both sides of the border and the critical
curve is the $x$-axis which has no cusp point, so cusps do not appear
on the invariant curve. Here the invariant curve is destroyed through
a homoclinic bifurcation. We have shown that contact bifurcation
occurs when the stable manifold touches the critical curve. This
bifurcation causes a qualitative change in the structure of basin of
attraction, from simply connected and unbounded to nonconnected and
bounded. At that point, the stable manifolds are self-intersecting
since the system has become structurally unstable after the appearance
of homoclinic tangle. We have observed the hysteretic transitions
between mode-locked dynamics to chaotic dynamics and vice versa and
have found the mechanism to be quite different from that in invertible
maps.

\section*{Acknowledgment}
One of the authors (Soma De) thanks the CSIR, India, for financial
support in the form of a Senior Research Fellowship.


\end{document}